\documentclass[11pt,a4paper]{article}
\usepackage{jheppub}
\usepackage{bbm}
\usepackage{amsmath}
\usepackage{stackrel}
\newcommand{\Ein}{E_\mathrm{in}}
\newcommand{\Eout}{E_\mathrm{out}}
\newcommand{\be}{\begin{equation}}
\newcommand{\ee}{\end{equation}}
\newcommand{\ba}{\begin{eqnarray}}
\newcommand{\ea}{\end{eqnarray}}
\renewcommand{\(}{\left(}
\renewcommand{\)}{\right)}

\newcommand{\lk}{\left[}
\newcommand{\rk}{\right]}

\newcommand{\w}{\omega}
\newcommand{\hw}{\hat{\omega}}
\newcommand{\nn}{\nonumber}
\newcommand{\g}{\gamma}
\newcommand{\hq}{\hat{q}}

\title{Probing the pattern of holographic thermalization with photons}

\author[a]{Dominik Steineder,}
\author[a,b]{Stefan A.~Stricker,}
\author[b]{Aleksi Vuorinen}

\affiliation[a]{Institute of Theoretical Physics, Technical University of Vienna,\\ Wiedner Hauptstr.~8-10, A-1040 Vienna, Austria}

\affiliation[b]{Faculty of Physics, University of Bielefeld,
D-33615 Bielefeld, Germany}

\emailAdd{steineder@hep.itp.tuwien.ac.at}
\emailAdd{stricker@hep.itp.tuwien.ac.at}
\emailAdd{vuorinen@physik.uni-bielefeld.de}

\abstract{We investigate the behavior of the retarded Green's function of a U(1) gauge field in holographic ${\mathcal N}=4$ Super Yang-Mills plasma, taking the leading strong coupling corrections into account. First, we use the thermal limit of this quantity to determine the flow of the photon quasinormal mode spectrum away from the infinite 't Hooft coupling limit, and after this specialize to a specific model of holographic thermalization, in which we evaluate the corresponding spectral density. In the latter case, our primary interest lies in the pattern, with which the spectral density approaches its equilibrium form, and how this process depends on the value of the coupling as well as the photon virtuality. All of the results obtained point consistently towards the weakening of the usual top/down pattern of holographic thermalization, once the coupling is decreased from the $\lambda=\infty$ limit.}

\keywords{AdS/CFT Correspondence, Quark Gluon Plasma}

\begin{document}

\rightline{BI-TP 2013/09}

\rightline{TUW-13-05}
\maketitle

\section{Introduction}

Quantitatively understanding the processes driving the complicated field dynamics during a relativistic heavy ion collision presents an exceptionally difficult and topical challenge to QCD theorists. According to standard lore, the early onset of hydrodynamical behavior as well as the apparently very low value of the shear viscosity of the produced matter, both inferred from RHIC and LHC data, point towards the system being a strongly coupled, nearly ideal fluid \cite{Tannenbaum:2012ma,Muller:2012zq}. This has led to gauge/gravity methods becoming standard tools in the description of quark gluon plasma (QGP) physics, filling in gaps left by standard approaches such as perturbation theory and lattice simulations (see e.g.~\cite{CasalderreySolana:2011us}). The success of the holographic approach relies largely on the fact that even though in their respective vacuum states QCD and conformal ${\mathcal N}=4$ Super Yang-Mills (SYM) theory are clearly very different, at high temperatures they both describe 
deconfined plasmas with qualitatively similar properties. 

A particularly useful recent development in the holographic description of QCD matter has been the mapping of the off-equilibrium dynamics driving thermalization in a strongly interacting plasma to the gravitational problem of black hole formation in asymptotically Anti de Sitter (AdS) spacetimes. Although this dual system is computationally almost equally challenging to approach, gauge/gravity calculations have already led to impressive results, such as a qualitative resolution of the famous fast apparent thermalization puzzle by demonstrating that the applicability of hydrodynamics does not require the system to be isotropic, let alone thermal \cite{Chesler:2008hg, Chesler:2010bi, Heller:2013fn}. Other recent applications of holography in this context include e.g.~progress in the evaluation of off-equilibrium Green's functions \cite{CaronHuot:2011dr}, developments in various thermalization scenarios (see e.g.~\cite{Balasubramanian:2010ce,Balasubramanian:2011ur,Galante:2012pv,
Erdmenger:2012xu}), as well as the determination of a host of kinetic theory parameters at strong coupling \cite{Kovtun:2004de,Rebhan:2011vd,Herzog:2006gh,CasalderreySolana:2006rq}.

A particularly important challenge in the study of the thermalization process in heavy ion collisions is to identify the pattern, with which plasma constituents with different energies approach their final thermal distributions. On the weak coupling side, a classical result identified elastic scatterings as the dominant process behind a bottom-up type thermalization \cite{Baier:2000sb}, which has recently received confirmation from classical gauge theory simulations \cite{Berges:2013eia} (see also \cite{Kurkela:2011ti} for a competing/complementing proposition emphasizing the role of plasma instabilities). In contrast, all holographic studies of thermalization in infinitely strongly coupled gauge theory have consistently pointed towards a top-down pattern, in which the modes with highest energy are the first to thermalize. This observation, which can be understood via a simple causal argument, indicates the probable existence of a transition between the two thermalization patterns at intermediate coupling. 
While the details of this conclusion certainly depend on the initial conditions the system is subjected to, it clearly poses an interesting topic for further study.

A convenient way of studying the thermalization pattern of a plasma with given initial conditions and coupling strength is through the determination of various off-equilibrium Green's functions, whose evolution gives direct information about the rate with which plasma constituents of different energy approach the thermal limit. Due to technical complications inherent in their holographic determination (see e.g.~\cite{CaronHuot:2011dr}), one is in practice forced to work within the simplest and most symmetric models of thermalization, such as one where the process is dual to the radial collapse of a thin spherical shell in AdS$_5$ space and the subsequent formation of a black hole. This model, originally proposed by \cite{Danielsson:1999fa, Danielsson:1999zt} and further studied e.g.~in \cite{Lin:2008rw}, corresponds to an initial state rather far away from the early stages of a real life heavy ion collision.  At the same time, it however allows for rather complicated calculations to be carried out, such as 
the determination of off-equilibrium spectral densities and particle production rates \cite{Lin:2008rw,Baier:2012tc, Baier:2012ax,Balasubramanian:2012tu}.

In the phenomenological description of the thermalizing plasma, one particularly interesting observable is the spectrum of emitted photons, as these particles interact with the plasma constituents so weakly that they escape the system almost unaltered. In thermal equilibrium, their production has been considered to leading as well as next-to-leading order (LO and NLO) in perturbative QCD \cite{Aurenche:1983ws,Ghiglieri:2013gia}, to LO in the limits of weakly and strongly coupled ${\mathcal N}=4$ SYM theory \cite{CaronHuot:2006te}, and in the latter theory even to NLO in a strong 't Hooft coupling expansion \cite{Hassanain:2011ce, Hassanain:2012uj}, including corrections of order $1/\lambda^{3/2}$. The LO holographic calculations have since then been generalized to an out-of-equilibrium setup \cite{Baier:2012tc, Baier:2012ax}, and later to NLO for the case of real photons \cite{Steineder:2012si}. These studies revealed interesting details of the virtuality and coupling strength dependence of the 
thermalization process, some of which were in fact rather striking. In particular, the NLO study of photon emission in \cite{Steineder:2012si} found indications of the top-down pattern of holographic thermalization shifting towards bottom-up type behavior, once the leading finite coupling corrections were taken into account.

In the paper at hand, our aim is to continue the study of holographic thermalization through a further analysis of photon production in ${\mathcal N}=4$ SYM plasma. In particular, we plan to fill a number of gaps left by our earlier works \cite{Baier:2012tc, Baier:2012ax,Steineder:2012si} by completing the following three tasks:
\begin{enumerate}
 \item Determine the quasinormal mode (QNM) spectrum of the photons from the corresponding retarded Green's function in thermal equilibrium, working at NLO in a strong coupling expansion. Study the flow of the QNM poles with $\lambda$ for several different virtualities.
 \item Compute the photon spectral density for different virtualities and couplings, limited only by the convergence of the strong coupling expansion.
 \item Perform an analytic WKB-type calculation for the asymptotic large-$\omega$ behavior of the photon spectral density, accounting for the ${\mathcal O}(1/\lambda^{3/2})$ corrections. 
\end{enumerate}
We anticipate that these three results will shed more light on the conclusions of \cite{Steineder:2012si}, either strengthening or challenging the claims made therein. Doing this, we also hope to identify the most fruitful directions of future research within these topics.

Our paper is organized as follows. Section 2 contains a brief introduction to the collapsing shell model of \cite{Danielsson:1999fa}, while section 3 lays out the general strategy, with which we determine the photon Green's functions. In both of these cases, we begin from the $\lambda=\infty$ limit, and in separate subsections consider the modifications introduced by including the leading ${\mathcal O}(1/\lambda^{3/2})$ corrections in the calculations. After this, section 4 is devoted to presenting the results for both the QNM spectra and the spectral densities, while in section 5 we draw our conclusions. Finally, the computational details related to the WKB calculation, discussed also in section 4, are left to appendix A.

\section{The setup}

\subsection{Generalities}

We wish to use the collapsing shell model of \cite{Danielsson:1999fa, Danielsson:1999zt} to gain insights into the thermalization process of strongly coupled $\mathcal{N}=4$ SYM plasma. On the dual gravity side, this model describes the gravitational collapse of a spherically symmetric, infinitely thin shell of matter in the radial direction of AdS$_5$ space --- a process dependent on the initial condition of the shell as well as its equation of state. Leaving these details unspecified, we parameterize the progress of the system by the radial location of the shell $r_s$, which eventually approaches a value corresponding to the Schwarzschild horizon of the forming black hole, $r_h$. On the field theory side, the setup may be thought of as resulting from an instantaneous, spatially homogeneous injection of energy into the vacuum state of the theory; for details of the realization of such a system, see e.g.~\cite{Wu:2012rib,Wu:2013qi,olli}

Throughout our calculations, we work in the quasistatic limit of the dynamics, in which the motion of the shell is considered slow in comparison with the other time scales of interest. As discussed in \cite{Baier:2012tc}, in the case of a shell composed of null dust (obeying the relation ${\rm d}s^2=0$), its approach towards the horizon takes the form
\ba
\frac{r_s-r_h}{r_h}&=& e^{-\frac{t}{\tau_h}}\, , \label{time}
\ea
where $t$ denotes the time coordinate on the boundary (and thus in the dual field theory) and $\tau_h\equiv 1/(4\pi T)$, with $T$ denoting the temperature of the field theory system in its final thermal state. The approximation of neglecting the inverse time scale associated with the motion of the shell in comparison with the frequencies (momenta) of the Green's functions of interest thus improves with time, and only breaks down in the limits of early times and very small energies.

Following Birkhoff's theorem, we know that outside the shell the metric of our system corresponds to that of an AdS black hole, whereas inside the shell we have pure AdS$_5$ space. Changing the radial coordinate to $u\equiv r_h^2/r^2$, this means the metric obtains the compact form 
\be\label{AdS5}
ds^2=\frac{r_h^2}{L^2 u}\Big( f(u)dt^2+dx^2+dy^2+dz^2\Big)+\frac{L^2}{4 u^2f(u)}du^2 \, ,
\ee
where
\be
f(u) \,\equiv\, \left\{ \begin{array}{lr}
f_+(u)=1- u^2 \;\; & \;\; \mathrm{for}\;u <u_s\\
f_-(u)=1 & \mathrm{for}\; u > u_s \, .
\end{array}\right. \, \label{fu}
\ee
Here and in the following, we have chosen the subscript `--' to denote the inside and `+' the outside of the shell. As discussed in \cite{Baier:2012tc}, the discontinuity of the time coordinate at the shell implies that frequencies measured inside and outside the shell must also differ, and are in fact related through
\ba\label{omega}
\omega_-&=&\frac{\omega_+}{\sqrt{f_m}}\equiv \frac{\omega}{\sqrt{f_m}} \, ,\quad
f_m \, \equiv\,f_+(u_s).
\ea
Note that consistently with this notation, $\omega$ will below always refer to a frequency measured outside the shell.

Finally, we note that a further issue resulting from the presence of the shell is the changing boundary conditions of various fields at $r=r_s$ or $u=u_s$. For an electric field $E$, a straightforward application of the Israel junction condition gives (for more details, see e.g.~\cite{Lin:2008rw,Baier:2012tc})
\be\label{mc}
\begin{aligned}
E_- (u,\omega_-)|_{u_s}&=&\sqrt{f_m}E_+(u,\omega_+)|_{u_s}\, ,\\
\partial_u E_- (u,\omega_-)|_{u_s}&=&f_m \partial_u E_+(u,\omega_+)|_{u_s}\, ,
\end{aligned}
\ee
which will be an important ingredient of our determination of the retarded Green's function of a photon in the following sections.

\subsection{Finite coupling corrections}

In order to perform calculations beyond the usual $\lambda=\infty$ limit of the AdS/CFT correspondence, we must account for all corrections of order $\alpha'^3$ in the inverse string tension. In particular, this requires supplementing the usual type IIB supergravity action by terms proportional to $\gamma\equiv\frac{1}{8}\zeta(3)\lambda^{-\frac{3}{2}}$,
\be
S_\text{IIB}=S_\text{IIB}^0+\gamma S_\text{IIB}^{1} +{\mathcal O}(\gamma^2) \,,
\ee
where the expansion is performed around the ${\mathcal O}(\gamma^0)$ action
\be
S_\text{IIB}^0=\frac{1}{2\kappa_{10}}\int d^{10}x\sqrt{-g}\lk R_{10}-\frac{1}{2}(\partial \phi)^2-\frac{1}{4.5!}(F_5)^2 \, ,\rk
\ee
and the correction term obtains the form \cite{Paulos:2008tn, Myers:2008yi}
\be\label{actioncorr}
S_\text{IIB}^{1}=\frac{L^6}{2 \kappa_{10}^2}\int d^{10}x\sqrt{-g} e^{\frac{-3}{2}}\phi \Big(C+\mathcal{T}\Big)^4 \, .
\ee
In these expressions, $R_{10}$ denotes the Ricci scalar, $\phi$ the dilaton field and $F_5$ the five-form field strength, while $C$ stands for the Weyl tensor and the tensor ${\mathcal T}$ is given by
\be
\mathcal{T}_{abcdef}=i\nabla_a F^+_{bcdef}+\frac{1}{16}\( F^+_{abcmn}F_{def}^{+\;\;mn}-3F^+_{abfmn}F_{dec}^{+\;\;mn}\) \, .
\ee
In the last equation, the index triplets $\{a,b,c\}$ and $\{d,e,f\}$ are understood to be first antisymmetrized with respect to all permutations, and the two triplets then symmetrized with respect to the interchange $abc\leftrightarrow def$. One should also note that our notation for the various contractions in eq.~(\ref{actioncorr}) is only schematic, and e.g.~the $C^4$ term in fact denotes the combination
\be
C_{hmnk}C_{pmnq}C_h^{\;rsp}C^q_{\;rsk}+\frac{1}{2}C^{hkmn}C_{pqmn}C_h^{\;rsp}C^q_{\;rsk} \, . \label{C4}
\ee 
For further details of this construction, see e.g.~\cite{Hassanain:2011ce}.

The derivation of the $\gamma$-corrected background metric from the above action is in principle a rather laborious task, which is however significantly simplified by the observation that the tensor $\mathcal{T}$ vanishes when evaluated with the ${\mathcal O}(\gamma^0)$ metric of eq.~(\ref{AdS5}) \cite{Myers:2008yi}. This implies that the $C^4$ term of eq.~(\ref{C4}) is the only one giving nonzero contributions to the ${\mathcal O}(\gamma)$ terms in the metric, eventually leading us to the result \cite{Gubser:1998nz,Pawelczyk:1998pb,Paulos:2008tn}
\ba\label{AdSg}
ds^2 = \frac{r_h^2}{u} \, \left(-f_+(u) \,
K^2(u) \, dt^2 + d\vec{x}^2\right) 
+ \frac{1}{4 u^2
f_+(u)} \, P^2(u) \, du^2 + L^2(u) \, d\Omega_5^2\, ,
\ea
where the various functions read
\ba
\!\! K(u) &=& e^{\gamma \, [a(u) + 4b(u)]}\,, \,\, P(u) = e^{\gamma \,
b(u)}\,, \,\, L(u) = e^{\gamma \, c(u)}\, , \nonumber \\
a(u) &=& -\frac{1625}{8} \, u^2 - 175 \, u^4 + \frac{10005}{16} \,
u^6 \, , \nonumber \\
b(u) &=& \frac{325}{8} \, u^2 + \frac{1075}{32} \, u^4
- \frac{4835}{32} \, u^6 \, , \nonumber \\
c(u) &=& \frac{15}{32} \, (1+u^2) \, u^4 \,.
\ea
The $\gamma$-corrected relation between $r_h$ and the field theory temperature finally obtains the form $r_h=\pi T/(1+\frac{265\g}{16})$.

\section{Photon production} \label{photons}

\subsection{Generalities}

Next, we wish to apply the falling shell setup to an investigation of photon production during the thermalization process, complementing the treatment of \cite{Baier:2012tc,Baier:2012ax} by keeping the virtuality of the photons $v=(\w^2- q^2)/\w^2$ arbitrary. To this end, we denote the photon four-momentum by $Q\equiv(\omega,\mathbf{q})$ and parameterized the magnitude of the three-momentum $\mathbf{q}$ by $q=c \w$, letting $c=\sqrt{1-v}$ vary from 0 to 1. These two extreme cases correspond respectively to dileptons at rest, considered in \cite{Baier:2012tc}, and to real photons, cf.~\cite{Baier:2012ax}. Again, we begin our treatment from the $\lambda=\infty$ limit, continuing only in the next subsection to the effects of ${\mathcal O}(\gamma)$ corrections.

In the context of ${\mathcal N}=4$ SYM, by `photon' we mean a gauge field coupled to a conserved current corresponding to one of the U(1) subgroups of the SU(4) R-symmetry of the theory, chosen so that all the ${\mathcal N}=4$ scalars and fermions obtain equal charges. The quantity we are after is then the retarded Green's function of this field, denoted by $\Pi_{\mu\nu}(Q)$, from the transverse and longitudinal components of which one obtains the trace of the corresponding spectral density,
\be
\chi^{\mu}_{\,\mu}(Q)=-4\, \mathrm{Im}\,\Pi_\perp(Q)-2\, \mathrm{Im}\,\Pi_\|(Q)\, . \label{trchi}
\ee 
Using the fact that the fluctuation dissipation theorem is known to hold in the quasistatic limit \cite{Baier:2012tc}, this allows us to evaluate the photon Wightman function
\be
\Pi^<_{\mu\nu}(Q)\,=\, n(\omega)\chi_{\mu\nu}(Q)\, , \quad n(\omega)\,=\,1/(e^{\omega\beta}-1)\, ,
\ee 
which is further related to the differential photon production rate
\be
\frac{d\Gamma}{d^4Q}=-\frac{\alpha \eta^{\mu\nu}\Pi^<_{\mu\nu}(Q)}{24 \pi^4 Q^2} \, , \label{prodrate}
\ee
with $\alpha$ denoting the fine structure constant.

The retarded Green's function is a convenient quantity for holographic calculations, as it can be evaluated with the (by now rather standard) methods developed in \cite{Son:2002sd}. To this end, we begin from the classical equation of motion (EoM) of an U(1) vector field in curved spacetime,
\be
\frac{1}{\sqrt{-g}}\partial_{\mu}[\sqrt{-g}g^{\mu\rho}g^{\nu\sigma}F_{\rho\sigma}]=0\, ,\qquad F_{\rho\sigma}=\partial_\rho A_\sigma-\partial_\sigma A_\rho \, ,
\ee
from which we obtain differential equations for the transverse and longitudinal components of the corresponding electric field. In the metric of eq.~(\ref{AdS5}), this produces the momentum space EoMs (see also \cite{Kovtun:2005ev, CaronHuot:2006te})
\ba
\label{eom}
E_\perp''(u)+\frac{f'(u)}{f(u)}E_\perp'(u)+\frac{\hw^2-\hat{q}^2 f(u)}{u f(u)^2}E_\perp(u)&=&0\\
E''_\|(u)+\frac{\hw^2 f'(u)}{f(u)(\hw^2-\hat{q}^2)}E'_\|(u)+\frac{\hw^2-\hat{q}^2f(u)}{u f(u)^2}E_\|(u)&=&0\, , \label{eom2}
\ea
where $f(u)$ denotes the function defined in eq.~(\ref{fu}) and where we have introduced the dimensionless parameters $\hw\equiv w/(2 \pi T)$, $\hat{q}\equiv q/(2 \pi T)$. Already from here, two special limits are clearly visible: For deeply virtual photons at $\hq=0$ (dileptons at rest) the two EoMs agree, while for real photons with $\hq=\hw$ only the transverse component contributes. In these two cases, it thus suffices to consider only eq.~(\ref{eom}).

Next, we proceed to solve the above equations, taking as an example the transverse electric field. Outside the shell, $u<u_s$, we have $f(u)=f_+(u)$, causing the equation (\ref{eom}) to have a singular points at the horizon, $u=1$, with the indicial equation
\be\label{indequ}
\alpha^2+\frac{\hw^2}{4}=0 \, .
\ee
The two solutions of this equation, $\alpha=\mp i\hw/2$, are readily identified as the infalling and outgoing modes, of which the former, corresponding to the minus sign, is the physical one in the thermal limit. In the presence of a falling shell, we must however keep both functions, and thus write the general `outside' solution as a linear combination of the infalling and outgoing modes
\be
E_+(u)=c_+ E_\text{in}(u)+c_- E_\text{out}(u) \, , 
\ee
with $c_\pm$ denoting unknown coefficients. The two functions appearing on the right hand side are finally solved numerically using the ans\"atze
\be
E_\text{in/out}(u)=(1-u)^{\mp\frac{i\hw}{2}}y_\text{in/out}(u)\, 
\ee
with the functions $y_\text{in/out}(u)$ satisfying the boundary conditions $y_\text{in/out}(u=1)=1$ and $y'_\text{in/out}(u=1)=0$.

Having obtained the electric field outside the shell (up to the coefficients $c_\pm$), the next step is to match it to the inside solution at $u=u_s$ using the matching conditions of eq.~(\ref{mc}). To this end, we note that for $u>u_s$, the transverse EoM becomes
\be
\label{eom0}
E_-''(u)+\frac{\hw^2-f_m\hat{q}^2}{f_m u}E_-(u)=0\, ,
\ee
where $f_m$ is the constant factor defined in eq.~(\ref{omega}), originating from the relation between the inside and outside frequencies. The solution of this equation is easily found and can be expressed in terms of Bessel functions in the form
\be
\label{E0}
E_-(u)=\sqrt{u}\( J_1\lk2\sqrt{u}\(\frac{\hw^2}{f_m}-\hat{q}^2\)\rk-iY_1\lk2\sqrt{u}\(\frac{\hw^2}{f_m}-\hat{q}^2\)\rk\)\, ,
\ee
which we plug into a relation for the ratio $c_-/c_+$ solved from the matching conditions, 
\begin{eqnarray}\label{Cmp}
\frac{c_-}{c_+} &=& -\,  \frac{\Ein\,\partial_u E_- - \sqrt{f_m} E_-\, \partial_u \Ein}
{\Eout\,\partial_u E_- - {\sqrt{f_m}} E_- \, \partial_r \Eout}\Bigg|_{u=u_s} \,.
\end{eqnarray}
With this parameter determined, we have found the behavior of the (transverse) electric field for all values of $u$ up to an irrelevant overall normalization constant.

Repeating the above procedure for the longitudinal component of the electric field, we obtain the retarded Green's function of the photon upon an application of the relation \cite{Son:2002sd}
\be
\Pi(\hw,\hq)=-\frac{N_c^2T^2}{8}\lim_{u\rightarrow 0}\frac{E'(u)}{E(u)}\,  \label{Pidef} 
\ee
separately to both the transverse and longitudinal components of $\Pi_{\mu\nu}(Q)$. Inserting the results to eq.~(\ref{trchi}), this gives us the trace of the corresponding spectral density $\chi^\mu_{\,\mu}(\hw,\hq,u_s)$ as well as its relative deviation from the equilibrium limit (denoted by `th' below),
\be
\label{R}
R(\hw,\hq,u_s)=\frac{\chi^\mu_{\,\mu}(\hw,\hq,u_s)-(\chi_\text{th})^{\mu}_{\mu}(\hw,\hq)}{(\chi_\text{th})^{\mu}_{\mu}(\hw,\hq)}\, .
\ee
The behavior of this quantity --- in particular the effect of the virtuality parameter $c$ on its $\omega$-dependence --- is one of the most central issues to be studied in the following sections of this paper.

\subsection{Finite coupling corrections}

To study photon production outside the infinite 't Hooft coupling limit in principle necessitates working out all the contractions in eq.~(\ref{actioncorr}) and subsequently determining the $\gamma$-corrected gauge field EoM. This rather tedious exercise can, however, be avoided in the case of a transverse vector field, for which all the necessary formulae are available in the literature \cite{Hassanain:2011ce, Hassanain:2009xw, Hassanain:2012uj}. Taking advantage of these results, we will now study the behavior of retarded correlators of a transverse electric field $E_\perp(u)$ in the falling shell setup, keeping track of all terms linear in $\gamma$. It is, however, good to recall that according to eqs.~(\ref{trchi})--(\ref{prodrate}), these results can be related to the photon production rate only in the limit of zero virtuality or $c=1$. 

Following the treatment of \cite{Hassanain:2011ce}, we begin by redifining the photon field according to 
\ba
\Psi(u) &\equiv& \Sigma(u) E_\perp(u)\, , \quad \Sigma(u)^{-1}\equiv1/\sqrt{f(u)}+\gamma p(u)\, , \label{sigmadef}
\ea
where the function $p(u)$ is given by
\ba
p(u)&=&\frac{u^2\big(11700 - u^2 \big[343897 -u(87539\,u +37760\,\hat{q}^2)\big]\big)}{288\sqrt{f(u)}} \, .
\ea
This reformulation has the advantage of reducing the system to a simple Schr\"odinger-type problem with the action 
\ba
S&=&-\frac{N_c^2r_h^2}{16\pi^2}\int_k \int \!{\rm d}u\, \bigg[\frac{1}{2}\Psi{\mathcal L}\Psi +\partial_u \Phi \bigg]\, ,
\ea
where the boundary term has the form $\Phi(u)\equiv \Psi'(u)\Psi(u)$ and the EoM for $\Psi(u)$ reads
\ba
\Psi''(u)&-&V(u)\Psi(u)=0\, , \label{EoM1} \\
V(u)&=&-\frac{1}{f^2}\bigg[\frac{u+\hw^2-\hq^2 f}{u}- \gamma\frac{f}{144} \Big(-11700 u + 2098482 u^3 - 4752055 u^5 \nonumber \\
&&+ 1838319 u^7+ 
 \hat{q}^2 (4770 + 11700 u^2 - 953781 u^4 + 1011173 u^6) \nonumber \\
 &&-  \hw^2(4770  + 28170 u^2 - 1199223 u^4)\Big) \bigg]\, . \label{V}
\ea

From this point on, solving for the retarded Green's function of the photon proceeds along the very same lines as in the $\lambda=\infty$ limit. Outside the shell, we write the ingoing and outgoing solutions to the EoM in the forms 
\be\label{psi}
\Psi_\text{in/out}(u)=(1-u)^{\mp\frac{i\hw}{2}}\(\psi^{(0)}_\text{in/out}(u)+\gamma\,\psi^{(1)}_\text{in/out}(u)+\mathcal{O}(\gamma^2)\)\, ,
\ee
then separate eq.~(\ref{EoM1}) into ${\mathcal O}(\gamma^0)$ and ${\mathcal O}(\gamma^1)$ parts, and finally solve $\Psi^{(0)}(u)$ and $\Psi^{(1)}(u)$ using the same boundary conditions as before. Just like at $\gamma=0$, this gives us the outside solution in terms of two unknown constants, 
\be
\Psi_+(u)=c_+ \Psi_\text{in}(u)+c_- \Psi_\text{out}(u) \, .
\ee
Inside the shell, we on the other hand use the fact that the metric (\ref{AdSg}) does not receive $\gamma$-corrections \cite{deHaro:2003zd, Banks:1998nr}, implying that the ${\mathcal O}(\gamma)$ part of the $\Psi_-$ field,
\be
\Psi_{-}(u)=\Psi_{-}^{(0)}(u)+\g \Psi_{-}^{(1)}(u)\, ,
\ee
originates solely from the relation (\ref{omega}) between the inside and outside frequencies, where we must simply change
\ba
f_m&\to& f_m^\gamma\,\equiv\, f_m K^2(u_s)\,.
\ea
This in particular means that $\Psi_{-}^{(0)}(u)$ can be read off directly from eq.~(10) of \cite{Baier:2012ax} (noting that inside the shell $\Psi(u)=E_\perp(u)$), while $\Psi_{-}^{(1)}(u)$ is available through a simple expansion of Bessel functions.

Moving on to patching the two solutions together, we note that when written in terms of the field $\Psi$, the matching conditions of eq.~(\ref{mc}) take the forms
\ba\label{mcg}
\Psi_{-}(u_s)&=&\sqrt{f_m^\gamma}\Psi_+(u_s)/\Sigma(u_s)\, , \\
\big(\partial_u \Psi_-(u)\big)|_{u=u_s}&=&f_m^\gamma \partial_u\big(\Psi_+(u)/\Sigma(u)\big)\big|_{u=u_s}\nonumber \, ,
\ea
from which one can derive the ${\mathcal O}(\gamma)$ equivalent of  eq.~(\ref{Cmp}). This leads to a $\gamma$-correction to the ratio $c_-/c_+$, with which the transverse component of the spectral density is available from the relation \cite{Hassanain:2011ce}
\be\label{spectral}
\chi_\perp(\hw,\hq,u_s,\gamma)=\frac{N_c^2 T^2}{2}\(1-\frac{265}{8}\g\)\mathrm{Im}\lk \frac{\Psi'_{+}(u)}{\Psi_{+}(u)}\rk \Bigg|_{u=0}\, ,
\ee
expanded to linear order in $\gamma$. This result is finally used to define the quantity
\be
\label{R2}
R_\perp(\hw,\hq,u_s,\gamma)=\frac{\chi_\perp(\hw,\hq,u_s,\gamma)-(\chi_\text{th})_\perp(\hw,\hq,\gamma)}{(\chi_\text{th})_\perp(\hw,\hq,\gamma)}\, ,
\ee
which serves as our ${\mathcal O}(\gamma)$ generalization of eq.~(\ref{R}) and the behavior of which we will thoroughly analyze in the next section. There we will also compare the large-$\omega$ limit of our numerical results to an analytic WKB expression derived for the $c=1$ spectral density in appendix A.

\section{Results}

After explaining the main steps of our computation above, we will now move on to a numerical investigation of the corresponding results. This will be performed in three parts. We begin from the limit of thermal equilibrium, and inspect the effect of infinitesimal perturbations on the system by solving the QNM spectrum from the thermal photon correlator and by analyzing the flow of the poles as functions of $\lambda$. Next, we move to the falling shell setup, and perform a study of the virtuality dependence of the out-of-equilibrium spectral density of the photon, first in the $\lambda=\infty$ limit and then taking the leading $\gamma$-corrections into account. Finally, we specialize to the UV ($\omega\to\infty$) limit of the ${\mathcal O}(\gamma)$ spectral density of real photons, and inspect how well the analytic WKB solution presented in appendix \ref{WKB} approximates our full numerical results.

\subsection{Quasinormal mode spectrum}\label{QNMs}

Quasinormal modes (in equilibrium) are the strong coupling equivalent of the quasiparticles encountered in weakly coupled thermal field theory, and characterize the response of the system to infinitesimal external perturbations. On the dual gravity side, their spectrum is obtained by considering linearized fluctuations of some bulk field with infalling boundary conditions at the black hole horizon, and solving for the poles of the corresponding retarded Green's function. The QNM spectrum has the generic form
\ba
\omega_n(q)&=&M_n(q)-i \Gamma_n(q),
\ea
where $q$ denotes the three-momentum of the mode, and the real quantities $M_n(q)$ and $\Gamma_n(q)$ correspond to the energy and width (decay rate) of the excitation.

\begin{figure}
\centering
\includegraphics[width=12cm]{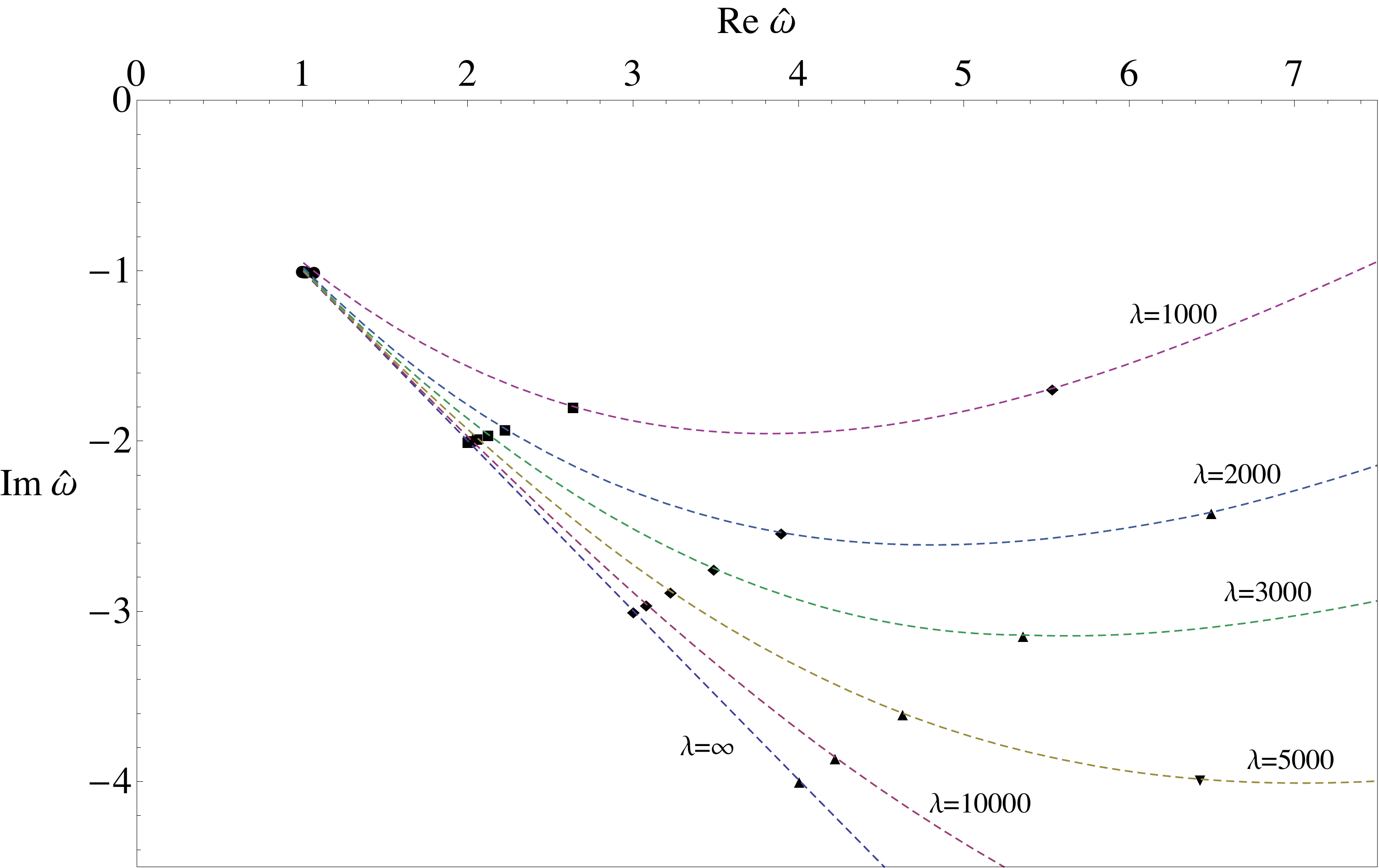}
\caption{The flow of the four lowest $q=0$ quasinormal poles on the complex $\omega$ plane as functions of $\lambda$. The first, second, third and fourth poles are denoted by circles, squares, diamonds and upward triangles, respectively, while the sole downward triangle corresponds to the fifth pole at $\lambda=5000$. The dashed lines of different colors correspond to 't Hooft coupling values ranging from $\lambda=\infty$ to $\lambda=1000$, and have been drawn here merely to quide the eye.}
\label{QNM1}
\end{figure}

For the case of the ${\mathcal N}=4$ photon correlator considered in this work, the QNM spectrum obtains a simple form in the respective limits of $q\to 0$ and $\lambda=\infty$ \cite{Kovtun:2005ev,CaronHuot:2006te},
\ba
\w_n(q=0)&=&2\pi T n(\pm1-i)\,,\qquad n=0,1,2,...
\ea
The fact that the widths of the excitations grow linearly as functions of their energy highlights the contrast between this system and one composed of longlived quasiparticles with Im$(\omega_n)\ll$ Re$(\omega_n)$. This observation can additionally be connected to the top-down nature of holographic thermalization at infinite coupling: Fluctuation modes with the highest energies are also the most shortlived ones, indicating that the hardest scales are the first ones to reach thermal equilibrium.

\begin{figure*}
\centering
\includegraphics[width=7.2cm]{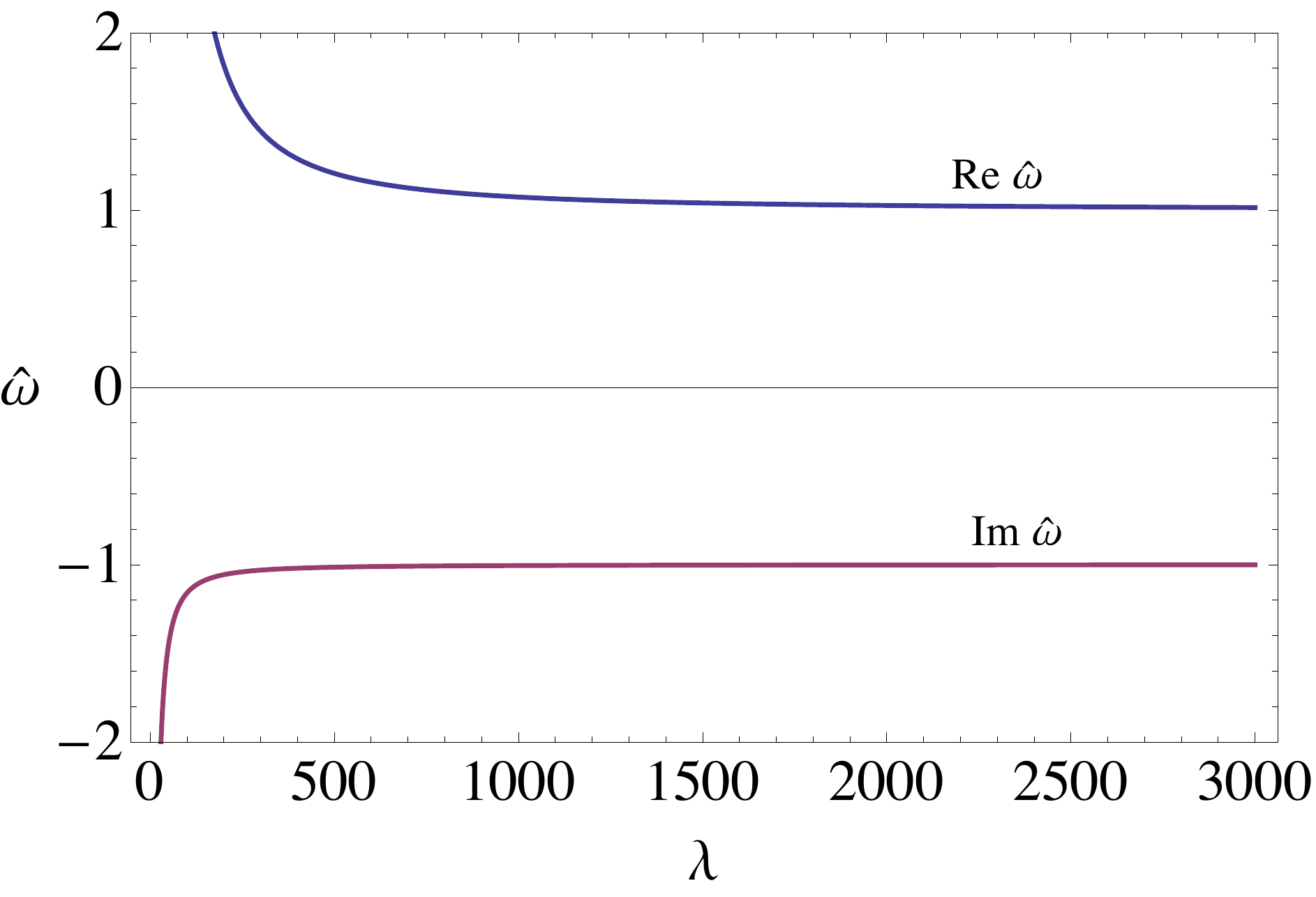}$\;\;\;\;\;\;\;\;$\includegraphics[width=7.2cm]{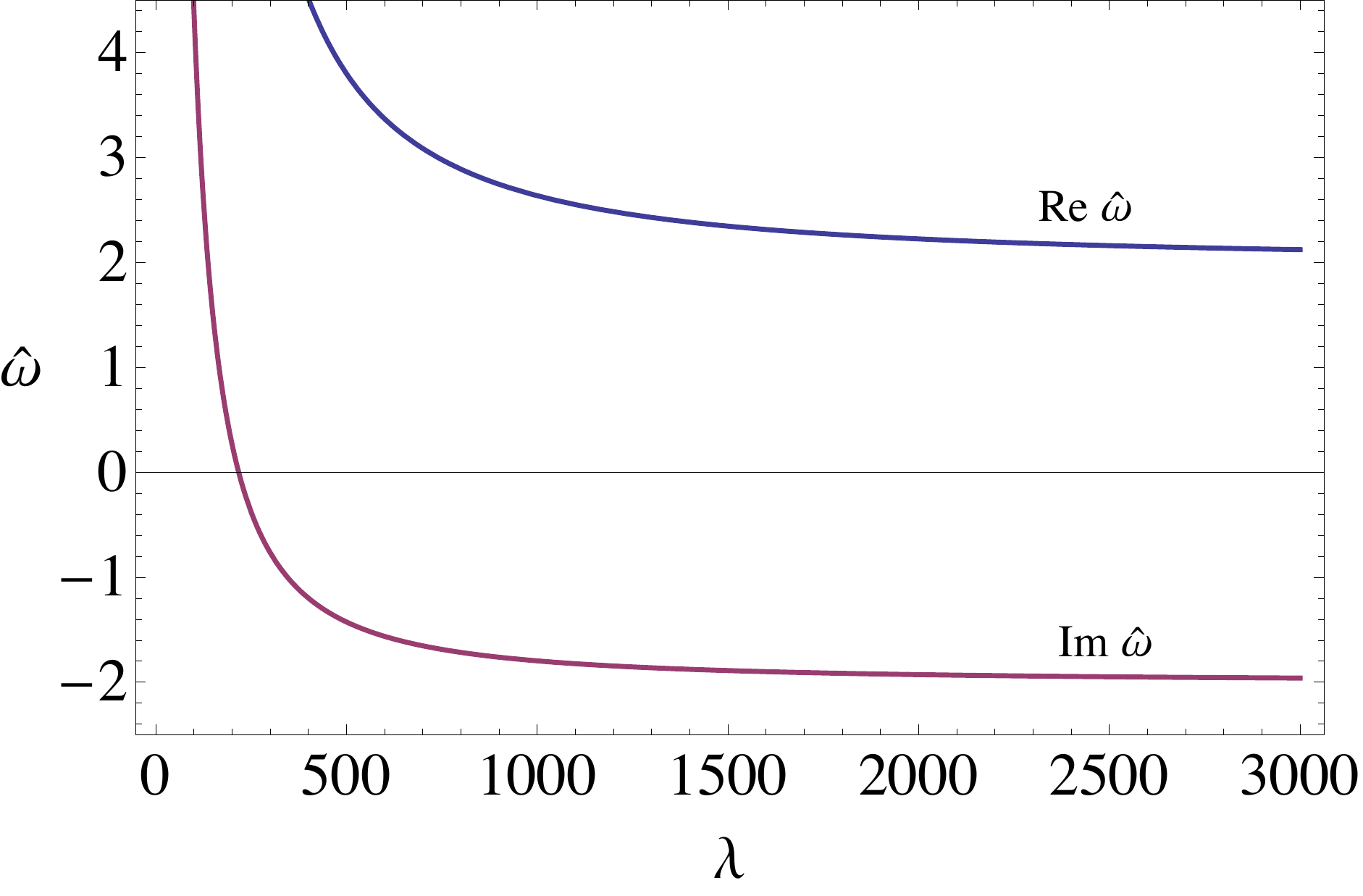}
\caption{The flow of the real and imaginary parts of the first  poles of fig.~\ref{QNM1} (circles and squares) as functions of $\lambda$.}
\label{QNM2}
\end{figure*}

In light of the above discussion, it is clearly very interesting to investigate, how the weak coupling quasiparticle spectrum turns into the QNM one, as the system becomes more strongly coupled. With the formalism introduced in the previous two sections, this problem can be approached from the strong coupling side by determining the poles of the thermal photon Green's function including its leading finite coupling corrections. Recalling eq.~(\ref{Pidef}), this calculation reduces to finding the zeros of the functions $\psi^{(0)}_\text{in}(u=0)$ and $\psi^{(1)}_\text{in}(u=0)$, introduced in eq.~(\ref{psi}), which we write in terms of Frobenius expansions \cite{Nunez:2003eq}
\ba
\psi_\text{in}^{(0)}(u,\omega)&=&\sum_{n=0}^N a_n(\omega) (1-u)^n\,, \quad
\psi_\text{in}^{(1)}(u,\omega)\,=\,\sum_{n=0}^N b_n(\omega) (1-u)^n\, .
\ea
The coefficients $a_n$ and $b_n$ appearing here are determined via an analytic solution of recursion relations provided by the EoM (\ref{EoM1}), while the parameter $N$ is chosen large enough so that the behavior of both functions becomes stable. The QNM spectrum is then obtained by making an ansatz for the frequency, $\omega_n = \omega_n^{(0)}+\gamma\omega_n^{(1)}$, substituting this to the equation $\psi_\text{in}^{(0)}(0,\omega)+\gamma\psi_\text{in}^{(1)}(0,\omega)=0$, and finally solving the system numerically order by order in $\gamma$.

\begin{figure}
\centering
\includegraphics[width=12cm]{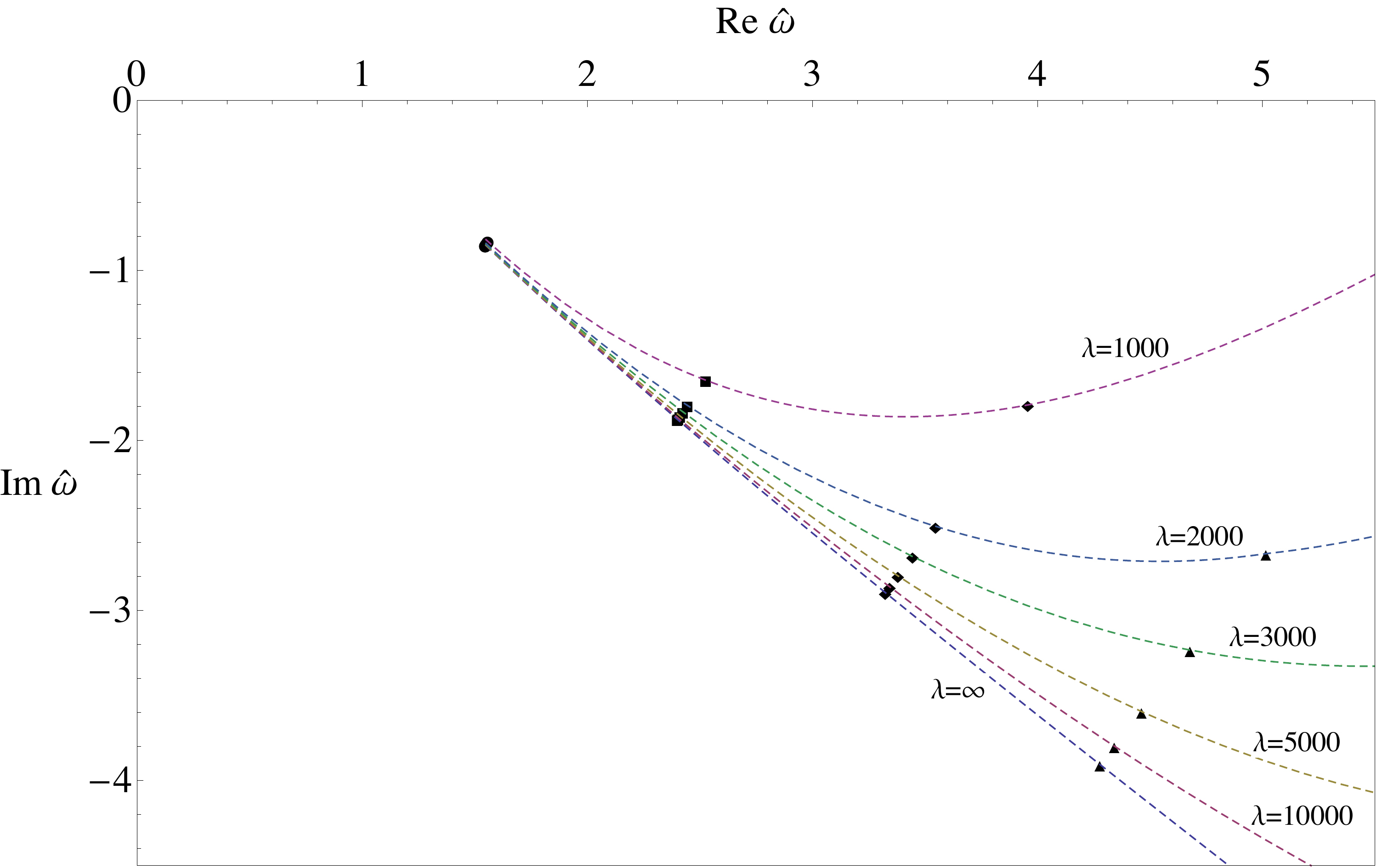}
\caption{Same as fig.~\ref{QNM1}, but for $\hat{q}=1$, i.e.~$q=2\pi T$.}
\label{QNM3}
\end{figure}

The results of the computation just described are displayed in figs.~\ref{QNM1}--\ref{QNM3}. In the first of these, corresponding to the $q=0$ case, we show the flow of the first four QNM poles on the complex $\omega$ plane, of which the first two are further inspected in fig.~\ref{QNM2}. As the 't Hooft coupling is decreased from the $\lambda=\infty$ limit towards $\lambda=1000$, we observe a clear bending of the spectrum away from the linear form. In particular, we witness a rapid increase of the imaginary parts of $\omega_n$ as $\lambda$ is lowered, as well as a strong dependence of the magnitude of the ${\mathcal O}(\gamma)$  term on the index $n$. These observations are suggestive of the QNM spectrum moving towards a quasiparticle one at smaller couplings, and are clearly consistent with the weakening of the top-down pattern of the thermalization process, cf.~\cite{Steineder:2012si}. It should be noted, though, that our strong coupling expansion can be trusted to converge only when the relative 
deviation of a pole location from its $\lambda=\infty$ value is small, which clearly is not the case for all of the poles shown in fig.~\ref{QNM1}. Finally, in fig.~\ref{QNM3} the same exercise is performed for the case of $\hat{q}=1$ (or $q=2\pi T$) with qualitatively similar results.

\subsection{Virtuality dependence of the spectral density} \label{virt}

\begin{figure*}
\centering
\includegraphics[width=7.5cm]{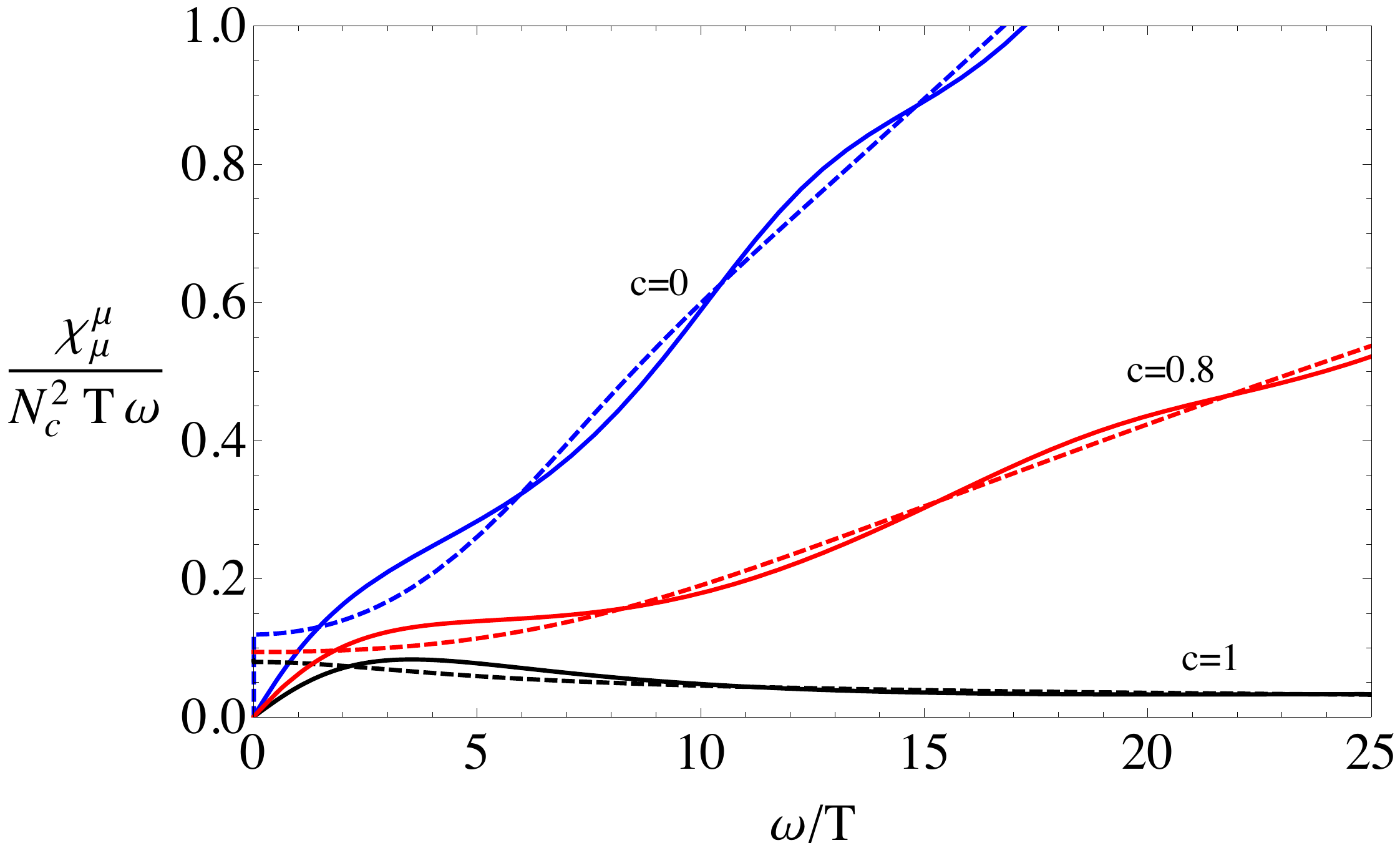}$\;\;\;\;\;\;\;\;$\includegraphics[width=7.0cm]{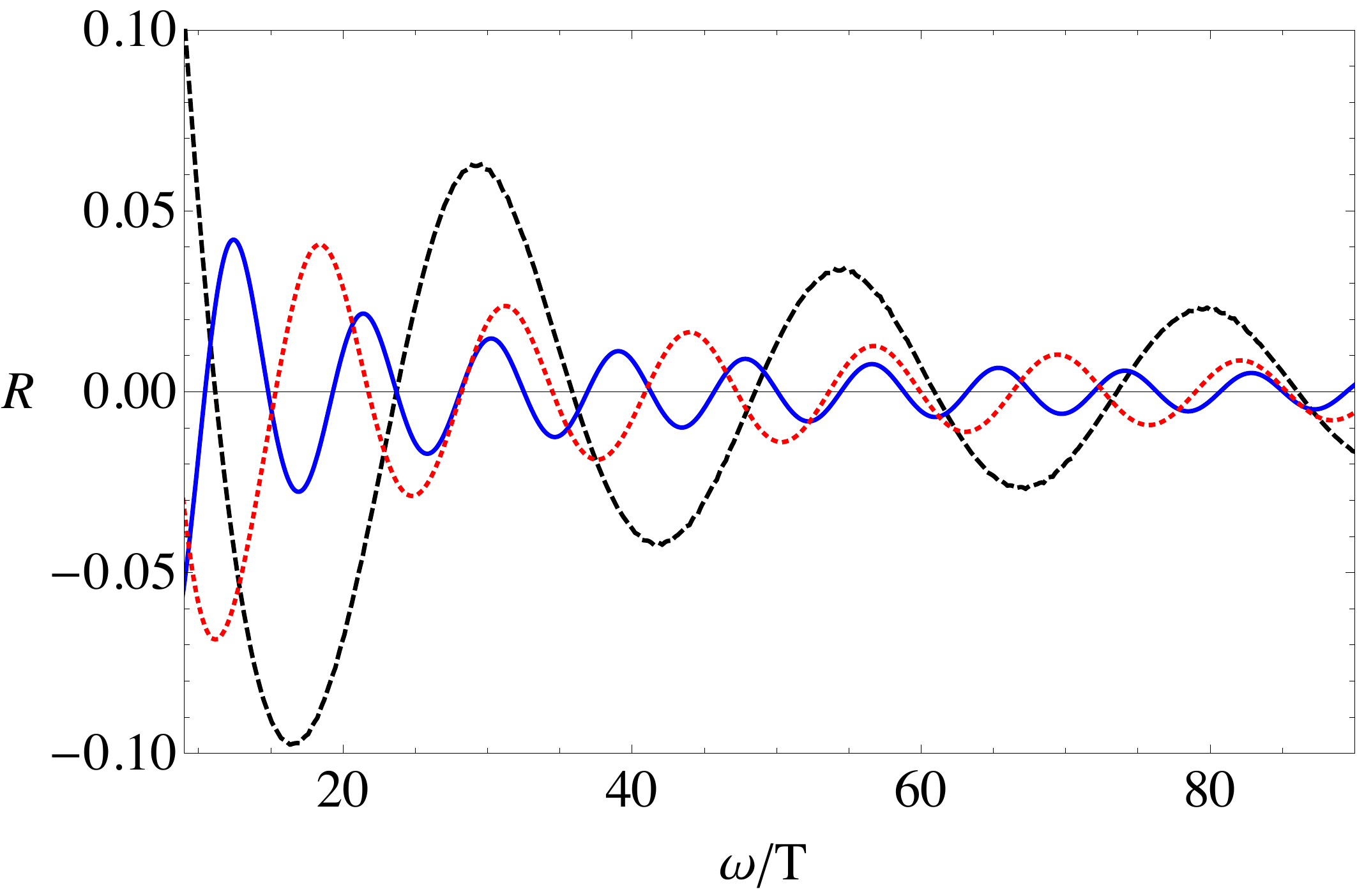}
\caption {Left: The $\lambda=\infty$ spectral density for photons of different virtualities in thermal equilibrium (dotted lines) and in an off-equilibrium state described by the collapsing shell model with $u_s=1/1.1^2$.  Right: The corresponding relative deviations from the thermal limit, $R$, for $c=0$ (solid blue line), $c=0.8$ (dotted red line) and $c=1$ (dashed black line).}
\label{VIR1}
\end{figure*}

Next, we move on to study the behavior of the retarded photon correlator in a specific out-of-equilibrium setting, namely during the thermalization process corresponding to the collapsing shell model. Beginning from the limit of $\lambda=\infty$, we want to complement our earlier work on real and maximally virtual photons \cite{Baier:2012tc, Baier:2012ax} by letting the virtuality $v=(\w^2- q^2)/\w^2$ be a free parameter. To this end, we parameterize $q=c\w$ and plot in fig.~\ref{VIR1} the behavior of the spectral density $\chi_\mu^\mu$ and its relative deviation from the thermal limit, $R$ (cf.~eq.~(\ref{R})), for $c=0$, 0.8 and 1. In analogy with the observations of \cite{Baier:2012tc, Baier:2012ax}, we again witness an oscillation of the off-equilibrium spectral densities around the corresponding equilibrium values, with the amplitude of the oscillations decreasing as the shell approaches the horizon radius.\footnote{Since the other effects of the shell location are very minor, we have in all of our plots 
set this parameter to the rather arbitrary value $u_s=1/1.1^2$.} The most important effect of the virtuality is clearly visible in the behavior of the relative deviation $R$: The larger the virtuality of the photon is (the smaller the value of $c$), the smaller is the amplitude of the oscillations. This implies that maximally virtual photons, i.e.~dileptons at rest, are the first ones to thermalize, confirming the observations made in somewhat different contexts in \cite{Arnold:2011qi, Chesler:2012zk}. Finally, we note that in all cases studied the amplitude of the oscillations in $R$ decreases at large $\omega$, consistent with the known top-down nature of thermalization at infinite 't Hooft coupling.

\begin{figure*}
\centering
\includegraphics[width=7.2cm]{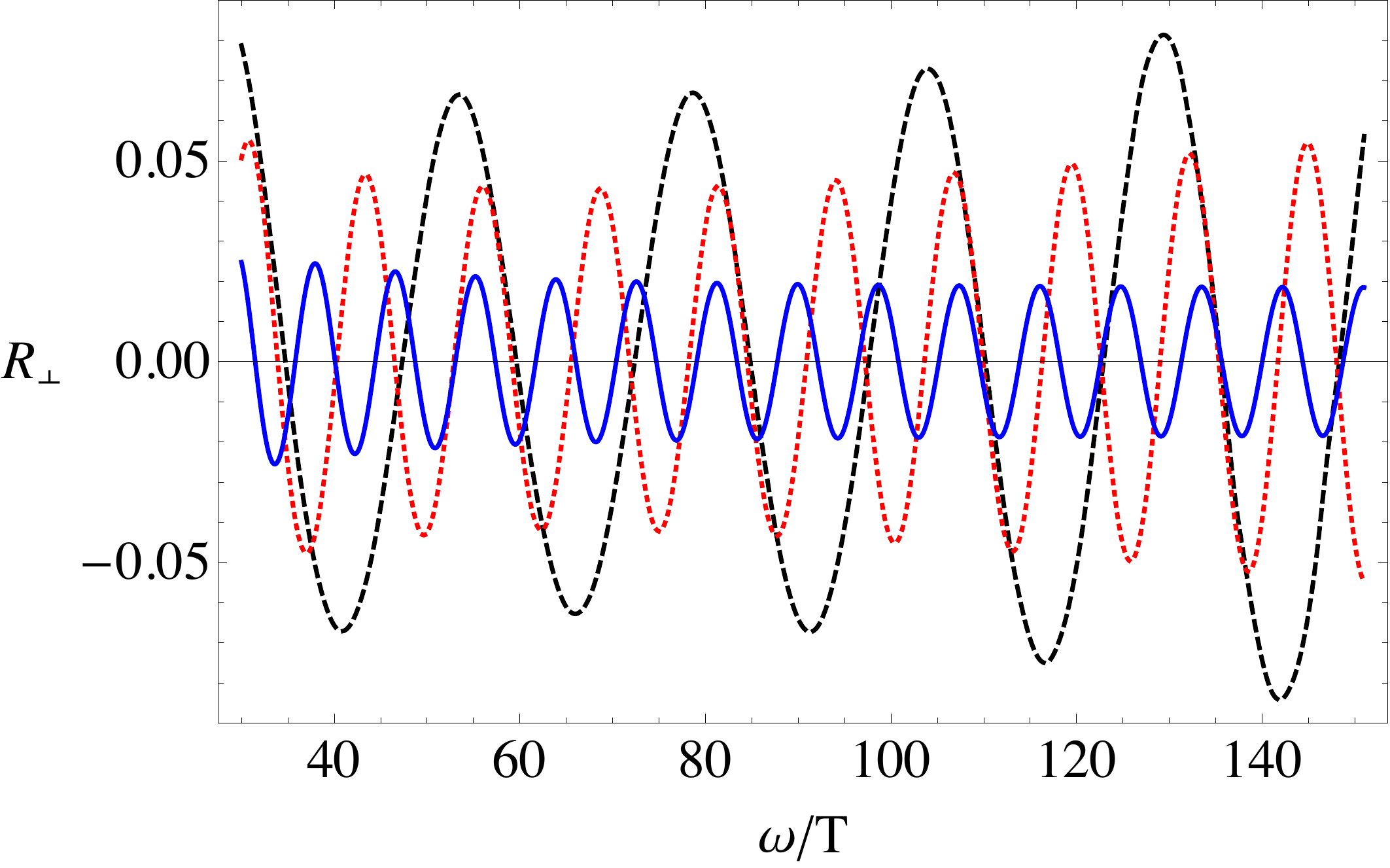}$\;\;\;\;\;\;\;\;$\includegraphics[width=7.2cm]{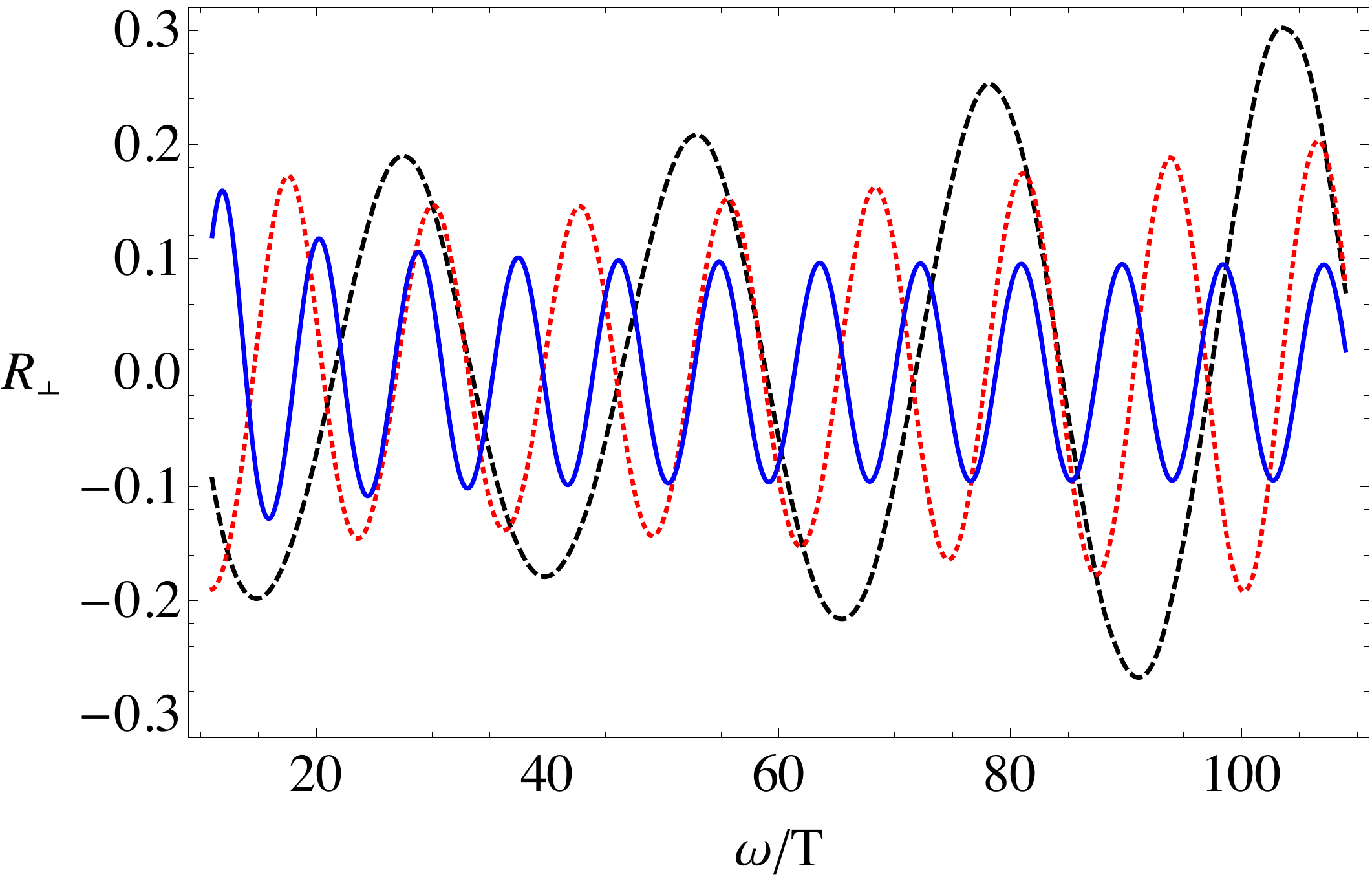}
\caption {As in fig.~\ref{VIR1}, but for $\lambda=300$ (left) and $\lambda=100$ (right).}
\label{VIR2}
\end{figure*}

At finite values of $\lambda$, one expects to witness a qualitative change in the behavior of the deviation function (now $R_\perp$), as reported for real photons ($c=1$) in \cite{Steineder:2012si}. To investigate what happens at nonzero virtuality, we determined the transverse spectral density $\chi_\perp$ and the corresponding $R_\perp$ for the same values of $c$ considered above, but setting now $\lambda=300$ and 100, cf.~fig.~\ref{VIR2}. Similarly to the case of real photons, we again observe the asymptotic behavior of the fluctuation amplitude changing from a $1/\omega$ suppression towards linear growth as the coupling is decreased. This behavior, however, appears to depend on the virtuality rather strongly, with the transition happening later as $c$ is decreased; in the case of maximal virtuality, $c=0$, the amplitude of $R_\perp$ even appears to approach a constant at large $\omega$. Finally, we note that the $\lambda=\infty$ observation of the amplitude of $R_\perp$ decreasing with increasing 
virtuality seems to hold at all coupling strengths considered.

Just as in the case of \cite{Steineder:2012si}, one must exercise some caution with the above results, not least because both the quasistatic approximation and the strong coupling expansion we have employed have finite regions of applicability. We note, however, that in the parameter ranges considered above, our results should be solid. First of all, a comparison of the relevant frequencies to the inverse time scale associated with the motion of the shell, cf.~eq.~(\ref{time}), clearly indicates that neglecting the latter is a reasonable approximation to make. And second, it can be explicitly checked that with the exception of the very largest frequencies studied, the deviation of $\chi_\perp$ from its $\lambda=\infty$ limit is always at most of the order of 10$\%$, giving us confidence that the strong coupling expansion is indeed well behaved.

\subsection{UV limit of the spectral density}\label{UV}

As the final test of our results, we want to study the extent one can understand the UV (large $\omega$) behavior of the spectral density of real photons analytically, recalling of course that in this formal limit the strong coupling expansion most likely breaks down. The calculation is nevertheless highly relevant for the arguments of \cite{Steineder:2012si} concerning the asymptotic behavior of $R$ being indicative of the thermalization pattern of the system.

\begin{figure}
\centering
\includegraphics[width=10cm]{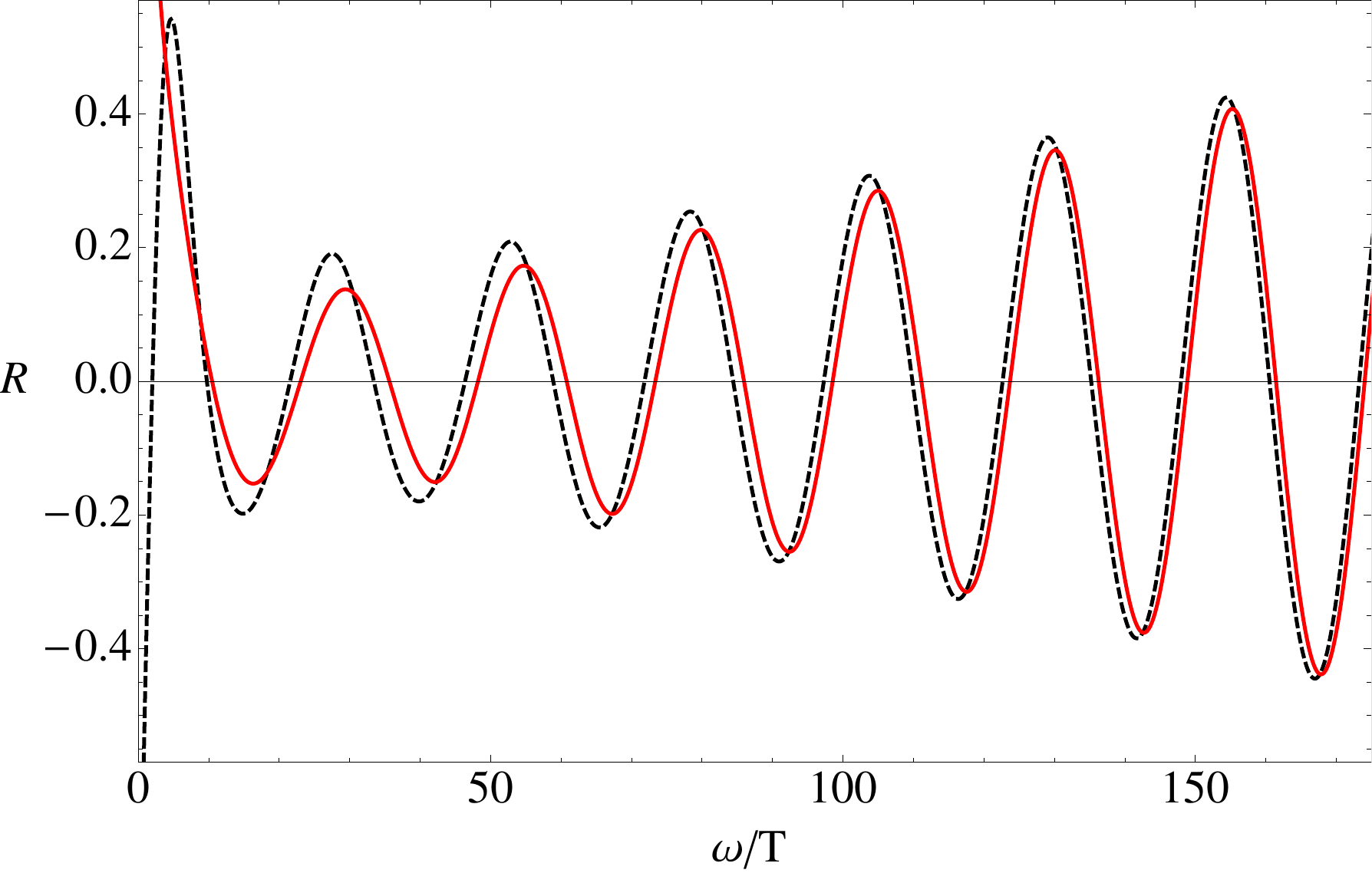}
\caption{A comparison of our exact numerical result for the relative deviation $R$ (dashed black line) with its analytic asymptotic limit derived from eq.~(\ref{WKBres}) (solid red line), with both results evaluated using $u_s=1/1.1^2$ and $\gamma=100$.}
\label{RWKB}
\end{figure}

In appendix \ref{WKB}, it is shown that a WKB-type calculation for the retarded photon correlator results in a relatively compact result for the large-$\omega$ behavior  of the ${\mathcal O}(\gamma)$ spectral density,
\ba
\chi_\text{asym}(\hat{\omega},u_s,\gamma)&=& \frac{N_c^2T^2}{4}\frac{3^{5/6}\Gamma(2/3)}{\Gamma(1/3)}\hat{\omega}^{2/3}
\Bigg\{1+5\gamma+{\mathcal O}(\gamma^2) \nonumber \\
&+&\frac{\sqrt{1-u_s^2}(1+\sqrt{1-u_s^2})}{8u_s^{3/2}}
\frac{\cos(\hat{\omega} g(u_s))+\sqrt{3}\sin(\hat{\omega} g(u_s))}{\hat{\omega}}+
{\mathcal O}(1/\hat{\omega}^2) \nonumber \\
&+&\frac{2950\gamma}{9}\bigg(u_s^{7/2}(1-u_s^2)\hat{\omega}\Big(\cos(\hat{\omega} g(u_s))+\sqrt{3}\sin(\hat{\omega} g(u_s))\Big)+{\mathcal O}(\hat{\omega}^0)\bigg) \nonumber \\
&+&{\mathcal O}(\gamma^2)\Bigg\}\, . \label{WKBres}
\ea
Here, one should note that while the order $\gamma^0$ off-equilibrium terms residing on the second line behave at large frequencies like $1/\omega$, the corresponding ${\mathcal O}(\gamma)$ terms on the third line indeed exhibit a linear behavior in $\omega$, consistent with our observations. In fig.~\ref{RWKB}, this comparison is made more quantitative by displaying the relative deviation $R$ obtained from this result with our full numerical calculation. As can be seen from the figure, the agreement between our numerics and the asymptotic large-$\omega$ limit is quite impressive already at moderate frequencies, improving consistently as one moves to even higher values of $\omega$.

\section{Discussion and conclusions}

\subsection{Summary of results}

In the paper at hand, we have addressed a number of open issues related to the production of (prompt) photons in a thermalizing holographic plasma, motivated by our earlier works on the same subject \cite{Baier:2012tc,Baier:2012ax,Steineder:2012si}. Our initial studies in the infinite 't Hooft coupling limit \cite{Baier:2012tc,Baier:2012ax} only covered the two limiting cases of real and maximally virtual photons ($c=1$ and $c=0$, respectively), and thus left open the determination of the full virtuality dependence of the results. More importantly, \cite{Steineder:2012si} made some highly nontrivial claims about the thermalization pattern of the holographic plasma shifting away from the usual top-down behavior, as the 't Hooft coupling is decreased from the $\lambda=\infty$ limit. This conclusion was mostly based on the UV behavior of the retarded photon Green's function, which was, however, studied only numerically. Moreover, all of these results were derived using only one particular model of holographic 
thermalization, and a clear need for a more universal analysis was identified.

In the following, we briefly summarize the findings we have made in the present work, and explain how we believe they shed light on the issues identified above:

\subsubsection*{Quasinormal mode analysis} 

In section \ref{QNMs}, we investigated the flow of the photon QNM spectrum, as the first strong coupling corrections were added to the $\lambda=\infty$ limit $\omega_n= n(\pm 1 -i)$, and the 't Hooft coupling was decreased. Our findings,  displayed in figs.~\ref{QNM1}--\ref{QNM3}, are strongly suggestive of the bending of the spectrum upwards on the complex $\omega$ plane. One must, however, naturally exercise some caution with the exact locations of the poles that have shifted an ${\mathcal O}(1)$ distance from their $\lambda=\infty$ values, as such a shift raises doubts on the convergence of the strong coupling expansion employed. 

If taken at face value, our QNM result is significant for two distinct reasons. First, it indicates a clear shift of the system towards one described by longlived quasiparticles, often encountered in weakly coupled thermal plasmas. The fact that the widths of the excitations are no longer proportional to their energies can also be interpreted as a sign of the weakening of the top-down thermalization pattern of the infinitely strongly coupled system, and thus strengthens the arguments made in \cite{Steineder:2012si}. Second, perhaps more importantly our QNM study is completely independent of the thermalization model used in our earlier works, and any approximations associated with it. This result is thus one of considerably more universal value than those relying on the collapsing shell setup.

\subsubsection*{Virtuality dependence}

In section \ref{virt}, we analyzed the virtuality dependence of the photon spectral density both in the strict $\lambda=\infty$ limit and including the leading ${\mathcal O}(\gamma)$ corrections. At infinite coupling, the results depicted in fig.~\ref{VIR1} confirmed the observations of \cite{Arnold:2011qi,Chesler:2012zk}, suggesting that highly virtual field modes are the first ones to thermalize. At the same time, fig.~\ref{VIR2} extended these results to finite coupling, revealing a strong virtuality dependence in the large-$\omega$ behavior of the ${\mathcal O}(\gamma)$ spectral density. In particular, with maximal virtuality the relative deviation of the spectral density from its thermal limit was seen to approach a constant in the $\omega\to\infty$ limit, in contrast with a linear rise observed for all nonzero values of $c$.

\subsubsection*{UV behavior of the spectral density}

Section \ref{UV}, as well as appendix \ref{WKB}, finally dealt with the analytic evaluation of the large-$\omega$ limit of the photon spectral density in the limit of zero virtuality but including strong coupling corrections. A WKB-type calculation performed in the appendix was seen to result in a rather compact expression for the formal UV limit, displayed in eq.~(\ref{WKBres}), which furthermore agreed with our full numerical result already at intermediate frequencies, cf.~fig.~\ref{RWKB}. This result clearly strengthens the conclusions made in \cite{Steineder:2012si}, as it confirms the linear increase of $R$ in the $\omega\to\infty$ limit. At the same time, it is nevertheless good to recall that due to the leading $\gamma$-corrections to $R$ being accompanied by a factor of $\omega$, this formal limit will never be reached within the validity of the strong coupling expansion.

\subsection{Future directions}

The most important qualitative outcome of the present paper is clearly the strengthening and confirmation of the most central conclusions of \cite{Steineder:2012si}. Both the generalization of our earlier calculations to nonzero virtuality and the analytic treatment of the large-$\omega$ limit of the photon spectral density point towards the weakening of the top-down thermalization pattern as the 't Hooft coupling is decreased from the $\lambda=\infty$ limit. At the same time, the determination of the photon QNM spectrum revealed compelling evidence for the flow of the system towards one described by quasiparticles, as the coupling is decreased. As discussed above, this pattern is furthermore in accordance with the weakening of the top-down thermalization behavior, as it points towards the relaxation of the correspondence between the energy and inverse lifetime of the field fluctuations once the value of the coupling is lowered.

While the results obtain in this paper are in many ways promising, much more remains to be done. On one hand, progress in the study of fully dynamical thermalization away from the infinite coupling limit would clearly be extremely valuable, generalizing results such as \cite{Erdmenger:2012xu,Heller:2011ju,Heller:2012km,Baron:2012fv,Chesler:2010bi, Chesler:2012zk,Wu:2011yd,vanderSchee:2012qj} to finite $\lambda$. At the same time, our current results on photon production should be straightforward to generalize e.g.~to the case of energy momentum tensor correlators, providing direct information on the thermalizing plasma constituents. Both of these topics are indeed amongst the issues we are currently working on.

\section*{Acknowledgments}
We are indebted to Janne Alanen, Rolf Baier, Michal Heller, Ville Ker\"anen, Esko Keski-Vakkuri, Elias Kiritsis, Hiromichi Nishimura, Anton Rebhan, and Olli Taanila for useful discussions. S.S.~A.V.~would in addition like to thank the organizers of the Holograv 2013 workshop, in particular Keijo Kajantie, for hospitality. The work of D.S.~was supported by the Austrian Science Foundation FWF, project P22114, S.S.~and A.V.~by the Sofja Kovalevskaja programme of the Alexander von Humboldt Foundation, and S.S.~additionally by the FWF START project Y435-N16.

\begin{appendix}

\section{WKB determination of the spectral density} \label{WKB}

In this appendix, we will perform an analytic WKB-type study of the large-$\omega$ behavior of the retarded Green's function of a real (transverse) photon field, keeping all terms to linear order in $\gamma$. This exercise is greatly simplified by the fact that the EoM (\ref{EoM1}) satisfies the conditions of Theorem 3.1 of \cite{Olver}, concerning the asymptotic behavior of the solutions of differential equations of a certain type. Applied to the computation at hand, this implies that just like in the $\gamma=0$ case studied in \cite{Baier:2012ax}, our general large-$\omega$ solution can be expressed in terms of a linear combination of Airy functions,
\ba
\hspace{-0.4cm} \Psi_+ (u)&=&h^{-1/4}(-u) \Big\{c_1\,{\rm Ai}(\hat{\omega}^{2/3}\zeta(-u))+c_2\,{\rm Bi}(\hat{\omega}^{2/3}\zeta(-u))\Big\}\, . \label{Eperp}
\ea
Here, the $c_i$ denote unknown coefficients (with no relation to the previous $c_\pm$), while the functions $h(-u)$ and $\zeta(-u)$ read
\ba
\label{zetay} \zeta(-u)&\equiv& e^{-i\pi/3} \bigg\{\frac{3\, g(u)}{4}\bigg\}^{2/3}\, , \\
g(u)&\equiv &2\,{\rm arctanh}(\sqrt{u})-2\,{\rm arctan}(\sqrt{u}) \nonumber \\
&&-\gamma\frac{u^{3/2}}{5544}\bigg(422730 - 2699862 u^2 -7078211 u^4\bigg)\, , \\
h(-u)&\equiv&-\frac{u}{(1-u^2)^2\zeta(-u)}\Bigg\{1+\gamma \frac{1-u^2}{144}\bigg(16470-245442u^2-1011173u^4\bigg)\Bigg\}\, .
\ea
Similarly to the corresponding $\gamma=0$ expressions of \cite{Baier:2012ax}, the function $g(u)$ is real and positive in the range $0\leq u \leq 1$ for all reasonable values of $\gamma$ (this ceases to be true for $\lambda\lesssim 4.2$, but such small values of the coupling are not considered in this work).

Following the steps explained in \cite{Baier:2012ax}, it is straightforward to verify that the (correctly normalized) linear combinations of the Airy functions satisfying infalling and outgoing boundary conditions at the horizon take the respective forms
\ba
\Psi_\text{in}(u)&=&\frac{h^{-1/4}(-u)}{\alpha\beta\beta_\text{in}}\,{\rm Ai}(\hat{\omega}^{2/3}\zeta(-u))\, ,  \\
\Psi_\text{out}(u)&=&\frac{\beta h^{-1/4}(-u)}{2\alpha\beta_\text{out}}\Big(i{\rm Ai}(\hat{\omega}^{2/3}\zeta(-u))+{\rm Bi}(\hat{\omega}^{2/3}\zeta(-u))\Big)\nn\, ,
\ea
where we have denoted
\ba
\alpha&\equiv& \frac{2^{\hat{\omega}/2}e^{-i\pi/4}}{2\sqrt{\pi}\hat{\omega}^{1/6}}\, ,\quad\quad
\beta \,\equiv \, 2^{i\hat{\omega}}e^{-i\pi\hat{\omega}/4}\, , \\
\beta_\text{in}&\equiv&1-\frac{\gamma}{11088}\bigg(9419333+9355343 i \hat{\omega}+1453760\hat{\omega}^2\bigg)\, ,\\
\beta_\text{out}&\equiv&1-\frac{\gamma}{11088}\bigg(9419333-9355343 i \hat{\omega}+1453760\hat{\omega}^2\bigg)\, .
\ea
Using the identities
\ba
e^{-i\pi/12}\,{\rm Ai}(e^{-i\pi/3}y)&\stackbin[y\to\infty]{}{\to}&\frac{e^{2i/3\, y^{3/2}}}{2\sqrt{\pi}y^{1/4}} +{\mathcal O}(y^{-7/4})\,,\\
e^{-i\pi/12}\,{\rm Bi}(e^{-i\pi/3}y)&\stackbin[y\to\infty]{}{\to}&\frac{2e^{-2i/3\, y^{3/2}}-i\, e^{2i/3\, y^{3/2}}}{2\sqrt{\pi}y^{1/4}} +{\mathcal O}(y^{-7/4})
\ea
for the asymptotic behavior of the Airy functions, it is then rather straightforward to derive the large-$\omega$ limit of the retarded Green's function,
\ba
\hspace{-0.9cm}\Pi_\perp^\text{asym}(\hat{\omega},u_s,\gamma)&=&-\frac{N_c^2T^2}{8}\frac{3^{1/3}\Gamma(2/3)}{\Gamma(1/3)}(1+5\gamma)
\frac{1-2^{2i\hat{\omega}}e^{-i\pi/2(\hat{\omega}+1/3)}\frac{\beta_\text{in}}{\beta_\text{out}}\frac{c_-}{c_+}}{1+2^{2i\hat{\omega}}e^{-i\pi/2(\hat{\omega}-1/3)}\frac{\beta_\text{in}}{\beta_\text{out}}\frac{c_-}{c_+}} (-\hat{\omega})^{2/3}\, , \label{Pitgen} 
\ea
where the ratio of $c_-$ and $c_+$ remains as the only unknown parameter.

In determining the large-$\omega$ limit of $c_-/c_+$, we finally take advantage of eq.~(\ref{Cmp}), to which we plug the above asymptotic forms of the infalling and outgoing field modes. After some algebra, this produces the result
\ba
c_-/c_+&=& C_0+\gamma C_1 + {\mathcal O}(\gamma^2)\, , \\
C_0&\stackbin[\hat{\omega}\to\infty]{}{\to}&\frac{i}{8}\, \quad {\rm with} 2^{-2i\hat{\omega}}u_s^{-3/2}\hat{\omega}^{-1}\sqrt{1-u_s^2}(1+\sqrt{1-u_s^2})e^{i\hat{\omega}(\pi/2+g_0(u_s))}\, ,\\
C_1&\stackbin[\hat{\omega}\to\infty]{}{\to}&\frac{2950i}{9}\, 2^{-2i\hat{\omega}}u_s^{7/2}(1-u_s^2)\hat{\omega}e^{i\hat{\omega}(\pi/2+g_0(u_s))}\, ,
\ea
where $g_0(u)\equiv g(u)|_{\gamma=0}$. Inserting these expressions to eq.~(\ref{Pitgen}), we then obtain an analytic, albeit quite complicated expression for the retarded correlator. Upon taking the imaginary part of this result and dropping all terms beyond linear order in $\gamma$, this function however simplifies considerably, producing the rather elegant ${\mathcal O}(\gamma)$ generalization of eq.~(35) of \cite{Baier:2012ax},
\ba
\chi_\text{asym}(\hat{\omega},u_s,\gamma)&=& \frac{N_c^2T^2}{4}\frac{3^{5/6}\Gamma(2/3)}{\Gamma(1/3)}\hat{\omega}^{2/3}
\Bigg\{1+5\gamma+{\mathcal O}(\gamma^2) \nonumber \\
&+&\frac{\sqrt{1-u_s^2}(1+\sqrt{1-u_s^2})}{8u_s^{3/2}}
\frac{\cos(\hat{\omega} g(u_s))+\sqrt{3}\sin(\hat{\omega} g(u_s))}{\hat{\omega}}+
{\mathcal O}(1/\hat{\omega}^2) \nonumber \\
&+&\frac{2950\gamma}{9}\bigg(u_s^{7/2}(1-u_s^2)\hat{\omega}\Big(\cos(\hat{\omega} g(u_s))+\sqrt{3}\sin(\hat{\omega} g(u_s))\Big)+{\mathcal O}(\hat{\omega}^0)\bigg) \nonumber \\
&+&{\mathcal O}(\gamma^2)\Bigg\}\, . 
\ea
In this expression, the first line represents the equilibrium limit of the result, while the second contains the ${\mathcal O}(\gamma^0)$ and the third the ${\mathcal O}(\gamma^1)$ non-equilibrium corrections to it.

\end{appendix}


\bibliographystyle{JHEP-2}

\bibliography{refs}

\end{document}